\documentclass{article}

\usepackage[emptyfooter,nonatbib]{neurips_2025}

\usepackage[utf8]{inputenc} 
\usepackage[T1]{fontenc}    
\usepackage{hyperref}       
\usepackage{url}            
\usepackage{booktabs}       
\usepackage{amsfonts}       
\usepackage{nicefrac}       
\usepackage{microtype}      
\usepackage{xcolor}         

\usepackage{listings}
\usepackage{graphicx}
\usepackage{array}
\usepackage{todonotes}

\title{Universal Reusability in Recommender Systems: The Case for Dataset- and Task-Independent Frameworks}

\author{%
  Tri Kurniawan Wijaya\\ 
  Huawei Ireland Research Centre \\
  Dublin, Ireland\\
  \texttt{tri.kurniawan.wijaya@huawei.com}
  \And
  Xinyang Shao\\ 
  Huawei Ireland Research Centre \\
  Dublin, Ireland\\
  \texttt{xinyang.shao@huawei-partners.com}
  \And
  Gonzalo Fiz Pontiveros\\ 
  Huawei Ireland Research Centre \\
  Dublin, Ireland\\
  \texttt{gonzalo.fiz.pontiveros@huawei.com}
  \And
  Edoardo D'Amico\\ 
  Huawei Ireland Research Centre \\
  Dublin, Ireland\\
  \texttt{edoardo.damico@huawei-partners.com}
}

\begin{document}

\maketitle

\begin{abstract}
Recommender systems are pivotal in delivering personalized experiences across industries, yet their adoption and scalability remain hindered by the need for extensive dataset- and task-specific configurations. 
Existing systems often require significant manual intervention, domain expertise, and engineering effort to adapt to new datasets or tasks, creating barriers to entry and limiting reusability. 

In contrast, recent advancements in large language models (LLMs) have demonstrated the transformative potential of reusable systems, where a single model can handle diverse tasks without significant reconfiguration. 
Inspired by this paradigm, we propose the \textit{Dataset- and Task-Independent Recommender System (DTIRS)}, a framework aimed at maximizing the reusability of recommender systems while minimizing barriers to entry. 
Unlike LLMs, which achieve task generalization directly, DTIRS focuses on eliminating the need to rebuild or reconfigure recommendation pipelines for every new dataset or task, even though models may still need retraining on new data.

By leveraging the novel \textit{Dataset Description Language (DsDL)}, 
DTIRS enables standardized dataset descriptions and explicit task definitions, allowing autonomous feature engineering, model selection, and optimization. 
This paper introduces the concept of DTIRS and establishes a roadmap for transitioning from \emph{Level-1 automation} (dataset-agnostic but task-specific systems) to \emph{Level-2 automation} (fully dataset- and task-independent systems). 
Achieving this paradigm would 
maximize code reusability and lower barriers to adoption.
We discuss key challenges, including the trade-offs between generalization and specialization, computational overhead, and scalability, while presenting DsDL as a foundational tool for this vision. 
This work aims to introduce a new paradigm in recommender system by 
rethinking their design for universal reusability and accessibility.
\end{abstract}

\section{Introduction}
\label{sec:intro}

Recommender systems play a crucial role in personalizing content across e-commerce, entertainment, and information retrieval applications. 
However, despite significant advances in machine learning and deep learning, 
the deployment and reusability of recommender systems remain limited by their dependence on dataset-specific and task-specific configurations. 
Current recommendation models typically require manual feature engineering, hyperparameter tuning, and task-specific optimization, 
leading to less reusable codes that might prevent both research and industrial adoption.

In contrast, recent breakthroughs in large language models (LLMs)
have demonstrated an alternative paradigm—one where a single pre-trained model can perform multiple tasks with minimal to no reconfiguration~\cite{achiam2023gpt, brown2020language, devlin2019bert, dubey2024llama, guo2025deepseek, kojima2022large, team2023gemini, wang2024sta}. 
While LLMs have revolutionized natural language processing (NLP) through task generalization, recommendation systems remain constrained by dataset and task dependencies. 
A recommender system trained for click-through rate (CTR) prediction might not be 
directly applicable to rating prediction or top-N recommendation, 
even when the underlying data representation is similar. 

To address this challenge, we propose the \emph{Dataset- and Task-Independent Recommender System (DTIRS)}, 
a paradigm shift that aims to reduce dataset- and task-specific reconfiguration in recommender systems. 
It is enabled by the \emph{Dataset Description Language (DsDL)}, 
a structured schema that allows models to infer dataset properties and task objectives automatically. 
Unlike LLMs, which achieve task generalization directly, DTIRS focuses on eliminating the need to rebuild or reconfigure recommendation pipelines for every new dataset or task, even though models may still need retraining on new data
(Section~\ref{sec:dtirs}).
Rather than requiring users to manually adapt the recommendation code or pipeline for every new dataset, we use DsDL to describe feature types, define task objectives, and autonomously configure the recommendation model, including feature preprocessing, model selection, and optimization
(Section~\ref{sec:dsdl}).

Furthermore, we introduce a categorization of recommendation tasks based on the mathematical structure of their prediction targets, leading to four fundamental task types:
binary prediction (e.g., CTR prediction, purchase intent),
real-valued prediction (e.g., rating prediction, demand forecasting),
ordered list prediction (e.g., top-N recommendation, personalized ranking),
unordered list (set) prediction (e.g., next-basket prediction, multi-label recommendation).
By structuring recommendation tasks in this way, DTIRS provides a mathematical foundation for task generalization, allowing one framework to handle multiple recommendation problems without manual reconfiguration.

The key contributions of this paper are as follows:
\begin{itemize}
    \item We introduce the \emph{Dataset- and Task-Independent Recommender System (DTIRS)} as a paradigm shift to maximize code reusability and minimize dataset-specific and task-specific dependencies in recommendation pipelines.
    \item We propose the \emph{Dataset Description Language (DsDL)}, a standardized schema that enables automated dataset interpretation and task definition, serving as the foundation for dataset- and task-independent recommendation models.
    \item We outline a roadmap for achieving full automation in recommender systems, transitioning from \emph{Level-1 automation} (dataset-independent but task-specific models) to \emph{Level-2 automation} (fully dataset- and task-independent recommendation models).
\end{itemize}


\noindent
We also make our resource website \url{https://dtirs.gitlab.io} 
publicly available,
where we regularly publish updates on DsDL, 
sample datasets, and DTIRS code sample whenever applicable.
We hope this new paradigm will benefit both practitioners and academics.

\section{Making the Case for Universal Reusability}
\label{sec:universal}

In our practical experience, we develop recommender systems for large-scale applications, including e-commerce, digital advertising, and rich media content platform. 
In these domains, recommender systems are deployed in dynamic environments where datasets change frequently, new recommendation tasks emerge, and real-world constraints demand efficient adaptation. 
From this perspective, a critical challenge in recommender systems is the lack of universal reusability—current systems are often tailored to specific datasets and tasks, 
requiring manual reconfiguration and domain expertise to adapt to new settings.

While recommender systems have advanced significantly in areas such as deep learning-based recommendation and automated hyperparameter tuning, 
their reusability and deployment might still be hindered by strong dependencies on dataset-specific and task-specific configurations. 
Unlike fields such as NLP, where LLMs can be applied to multiple tasks with minimal reconfiguration, recommendation models are typically designed with a rigid structure, making transfer across datasets and tasks non-trivial.

In the following, we discusses the practical 
limitations of current recommender systems, 
examines why reusability remains a challenge, and explores how a dataset- and task-independent 
framework might improve reusability and accessibility in recommender system development.

\vspace{0.3cm}
\noindent\textbf{Challenges in Recommender System Reusability.}
Recommender systems typically follow a pipeline that involves dataset preprocessing, 
feature engineering, model development and model (hyperparameter) tuning or optimization. 
However, this pipeline is rarely transferable across datasets or tasks without significant modifications. 
Some key challenges include:
\begin{itemize}
    \item \emph{Dataset-Specific Feature Engineering}: Many recommender models rely on handcrafted feature transformations that are specific to a given dataset. The same model architecture might not perform well when applied to a different dataset with different feature distributions.
    \item \emph{Task-Specific Optimization}: 
    A model designed for click-through rate (CTR) prediction, 
    for example, would not be directly applicable to top-N recommendation or 
    next-basket prediction without changes in objective functions, loss functions, or evaluation metrics.
    \item \emph{Hyperparameter Tuning for Each Dataset}: 
    Model hyperparameters often require tuning per dataset, 
    making deployment across multiple datasets computationally expensive.
\end{itemize}
While recommender systems are highly effective in dataset-specific applications, their adaptability across datasets and tasks remains an open challenge.

\vspace{0.3cm}
\noindent\textbf{Implications for Practical Applications.}
From a practical perspective, the lack of universal reusability in recommender systems has several implications:
\begin{itemize}
    \item \emph{Barrier to Adoption for Non-Experts}: Since current recommender systems require significant domain expertise for setup and fine-tuning, smaller organizations and non-specialists might struggle to deploy effective models.
    \item \emph{Limited Research Reproducibility}: Many published recommendation models are optimized for specific datasets, making it difficult for researchers to validate and 
    generalize results~\cite{cavenaghi2023systematic, ferrari2019we, hidasi2023effect, werner2024reproduce, wijaya2025dataset}, 
    which also motivates the inception of the dedicated \emph{Reproducibility} tracks in top information retrieval venues. 
    Several frameworks are recently proposed to mitigate this problem~\cite{anelli2021elliot, michiels2022recpack, salah2020cornac, shao2024rboard, sun2020are, zhao2021recbole, zhu2022bars}.
    \item \emph{Deployment Costs}: Companies must invest significant engineering effort in customizing models for different datasets, which increases time-to-market.
\end{itemize}


\vspace{0.3cm}
\noindent\textbf{A Dataset- and Task-Independent Recommender System.}
This paper presents an alternative paradigm for recommender systems: a dataset- and task-independent recommender system (DTIRS). Instead of training a universal model, DTIRS focuses on creating a framework where recommendation models can be applied to new datasets and tasks with minimal manual reconfiguration. 
The key components of DTIRS include the Dataset Description Language (DsDL), a standardized schema that allows models to interpret dataset structure and task objectives dynamically.

\section{Related Work}
\label{sec:relwork}
Recommender systems have traditionally required dataset-specific and task-specific configurations, 
reducing their reusability. 
Recent advancements in automation, meta-learning, and neural architecture search (NAS) 
have attempted to address some of these challenges. 
Additionally, the success of foundational models in NLP 
provides valuable insights into building more generalizable recommender system frameworks. 

\subsection{Traditional Recommender Systems}
Traditional recommender systems typically fall into three main categories: 
collaborative filtering, 
content-based filtering, and 
hybrid models~\cite{adomavicius2005toward, balabanovic1997fab, koren2021advances, sarwar2001item, wu2023mmgef}.

\begin{itemize}
    \item \emph{Collaborative Filtering (CF)} leverages historical user-item interactions to predict future preferences. 
    Matrix factorization-based methods like SVD \cite{koren2009matrix} and ALS \cite{hu2008collaborative} 
    have been widely used but require dataset-specific tuning.
    \item \emph{Content-Based Filtering} recommends items based on feature similarities between user preferences and available content \cite{lops2011content}. These approaches rely on domain-specific feature extraction and manual configuration.
    \item \emph{Hybrid Models} combine CF and content-based approaches to improve accuracy \cite{burke2002hybrid}. While effective, hybrid models introduce additional complexity and require dataset-specific hyperparameter tuning.
\end{itemize}

Most traditional recommender systems are dataset-specific and task-specific, 
requiring human intervention for feature engineering, 
model selection, and hyperparameter tuning. 

\subsection{Automated Recommendation Systems}
Several approaches have attempted to automate aspects of recommender systems.

AutoRec introduced an autoencoder-based approach to automate recommendation~\cite{sedhain2015autorec}. 
It focused on optimizing collaborative filtering tasks and did not generalize across datasets or tasks.
Auto-Surprise~\cite{anand2020auto} and yet another AutoRec~\cite{wang2020autorec} automate both model search and hyperparameter tuning for deep recommendation models, supporting tasks such as rating prediction and click-through rate (CTR) prediction. 
LensKit-Auto~\cite{vente2023introducing} 
allows users to automatically select, optimize, and 
ensemble LensKit algorithms~\cite{ekstrand2011lenskit}; it also highlights the significant impact of hyperparameter optimization. 
Other platforms also exist to help researchers and practitioners in gluing together several stages of 
experiment configuration, output management, and post-processing, such as 
Elliot~\cite{anelli2021elliot},
RecPack~\cite{michiels2022recpack},
librec~\cite{sonboli2021librec}, 
or RecBole~\cite{zhao2021recbole}.
%
Similarly, however, the models they produced remain either dataset- or task-specific.

There have been several line of work using meta learning to utilize knowledge acquired 
from one user to another to mitigate the cold-start 
problem~\cite{lee2019melu, liu2022online, vartak2017meta, wu2023m2eu}.
While meta learning holds promise for improving model adaptability, current approaches still necessitate dataset-specific feature processing and task-specific model selection.
%


Neural Architecture Search (NAS) automates the discovery of optimal model architectures for recommendation \cite{elsken2019neural, liu2019darts, zoph2017neural}. 
Recent applications of NAS in RecSys include:
NAS-driven architecture search for personalized recommendations~\cite{jiang2024automatic, yu2023ihas, zhang2024dns, zhu2022ctr} 
%
and
utilizing neural search for optimizing embedding layers in 
large-scale recommender systems~\cite{chang2024pefa,joglekar2020neural}
While NAS reduces manual architecture design, it does not address
dataset generalization or task-independent learning.

\subsection{Large Language Models}


One notable advancements in AI has been the emergence of
Large Language Models (LLMs), such as BERT~\cite{devlin2019bert}, 
GPT-4~\cite{achiam2023gpt},
Gemini~\cite{team2023gemini},
Llama-3~\cite{dubey2024llama}, and
DeepSeek-R1~\cite{guo2025deepseek}. 
These models exhibit generalization capabilities
across multiple NLP tasks with minimal task-specific fine-tuning.
%
%
LLMs, trained on diverse and large-scale datasets, offer a hint 
for universal reusability. 
While the recommender systems community is still in the early stages of 
developing similar foundational models, 
recent efforts~\cite{jiang2023self, jiang2023knowledge, wang2024pre, yadav2024extreme} 
indicate a growing interest in this direction.


Transfer learning has played a transformative role in NLP 
by facilitating cross-task generalization. 
In recommender systems, 
transfer learning has shown promise in cross- and multi-domain 
scenarios~\cite{ariza2023exploiting, guan2022cross, hou2024ecat, zhang2023collaborative}. 
Continued research in this area holds potential for broader, 
more flexible applicability across datasets and tasks.


LLMs further contribute to this vision by enabling new paradigms in recommender systems, 
either as components in hybrid 
architectures~\cite{muhamed2021ctr, qu2024elephant, sun2024large, wei2024llmrec, wijaya2024rs4rs, wu2021empowering, xi2024towards} or as 
standalone recommenders~\cite{bao2023tallrec, dai2023uncovering, li2023prompt, lian2024recai, liao2024llara, kim2024large}. 
These efforts shows the potential of LLMs as one of the building blocks in realizing DTIRS 
and maximizing the reusability in recommender systems.

\section{DTIRS Formulation}
\label{sec:dtirs}

\begin{figure*}[ht]
    \centering
    \includegraphics[width=0.9\linewidth]{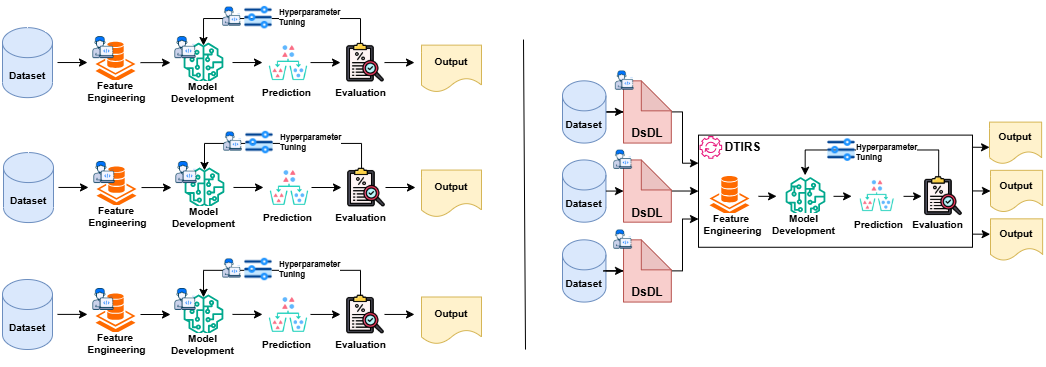}
    \caption{\small{Overview of the typical recommender system workflow (left) compared to our proposed DTIRS (right). The typical workflow often requires human expert and manual effort for feature engineering, model development, and hyperparameter tuning across different datasets, creating a barrier to entry; 
    the results are typically dataset- or task-specific codes or pipelines, reducing reusability. 
    In contrast, with the help of DsDL (Section~\ref{sec:dsdl}), 
    DTIRS aims to eliminate the need for manual reconfiguration in many of these steps, lowering the barrier to entry and increasing reusability.}}
    \label{fig:dtirs}
\end{figure*}

The Dataset- and Task-Independent Recommender System (DTIRS) aims to generalize recommendation models across different datasets and tasks without requiring manual reconfiguration. 
In this section, we formalize it, defining how DTIRS achieves dataset-independent and task-independent adaptability.
Figure~\ref{fig:dtirs} illustrates the comparison between the typical recommender system workflow and DTIRS.

\subsection{Problem Definition}

A standard recommender system takes an input dataset \( \mathcal{D} \) containing a set of features \( X \) and labels \( Y \), learning a function \( f \) such that:

\begin{equation}
    Y = f(X; \theta)
\end{equation}
where:
\begin{itemize}
    \item \( X \) represents the input feature space, including user, item, and contextual features.
    \item \( Y \) is the output, which varies depending on the recommendation task (e.g., binary classification, ranking, regression).
    \item \( \theta \) represents the model parameters, optimized during training.
\end{itemize}

Typical recommender systems require dataset-specific feature engineering and task-specific model selection, limiting their reusability. DTIRS seeks to eliminate these constraints by making \( f \) adaptable across datasets and tasks.

\subsection{Dataset-Independent Representation}

Instead of treating datasets as rigid structures, we represents a dataset $\mathcal{D}$ using a structured schema 
$S$ , provided via the Dataset Description Language (DsDL):

\begin{equation}
    S = (X, Y, \mathcal{T})
\end{equation}

\noindent
where:
\begin{itemize}
    \item \( X = \{ x_1, x_2, ..., x_m \} \) is the set of input features,
    \item \( Y \) is the output label, and
    \item \( \mathcal{T} \) is the task descriptor; it defines how \( Y \) should be predicted.
\end{itemize}

Given a new dataset $\mathcal{D}$, and its $S$, DTIRS dynamically configures the recommendation model accordingly.

\subsection{Automated Feature Engineering}
A key component of DTIRS is its ability to automatically preprocess and select features 
based on the dataset schema $S$. 
Typical recommender systems require manual feature engineering, 
which limits reusability across datasets. 
DTIRS attempts to eliminate this constraint.

DTIRS applies $\Phi(\cdot)$ that automatically processes input features based on their type and dataset schema:
\begin{equation}
    X' = \Phi(S)
\end{equation}
where $X'$ is the new feature set and $\Phi(\cdot)$ 
is a function or a composition of functions responsible for feature creation, transformation, and selection. 
In practice, these operations encompass, but are not limited to, 
handling missing values, encoding categorical features, 
normalization, scaling, text processing, list processing, 
variance filtering, correlation filtering, 
feature selection techniques such as backward or forward selection, 
as well as automated 
techniques~\cite{chen2019neural, horn2020autofeat, liu2021automated, wei2023automatic, zhu2022difer}.

\subsection{Task-Independent Modeling}

\begin{table*}[t]
\centering
\caption{Recommendation Tasks Categorization}
\label{tab:tasks}
\small{
\begin{tabular}{m{1.9cm} l m{2cm} m{4cm}}
\toprule
\textbf{Prediction Task Type} & \textbf{Output Structure} & \textbf{Typical Loss Function} & \textbf{Typical Evaluation Metrics} \\
\midrule
\midrule
Binary & $Y \in \{0,1\}$ & Binary Cross-Entropy & AUC, Precision, Recall, F1-Score, Log Loss \\
Real-Valued & $Y \in \mathbb{R}$ & MSE, MAE & RMSE, MAPE, $R^2$ \\
Ordered List & $Y = (y_1, y_2, ..., y_N), y_i \succ y_{i+1}$ & Pairwise loss, Listwise Loss & NDCG, MAP, MRR \\
Unordered List & $Y = \{y_1, y_2, ..., y_k\}$ & Jaccard Loss & Jaccard distance, Precision@K, Recall@K \\
\midrule
\bottomrule
\end{tabular}
}
\end{table*}

DTIRS generalizes across different recommendation tasks by categorizing 
tasks based on the mathematical structure of their outputs. 
We define four fundamental recommendation task types:
\begin{itemize}
    \item binary prediction (classification): \( Y \in \{0,1\} \) (e.g., click prediction),
    \item real-valued or numerical prediction (regression): \( Y \in \mathbb{R} \) (e.g., rating prediction),
    \item ordered list prediction (ranking): \( Y = (y_1, y_2, ..., y_k) \), where \( y_1 \succ y_2 \succ ... \succ y_N \) (e.g., Top-N recommendation), and
    \item unordered list prediction (set prediction): \( Y = \{y_1, y_2, ..., y_k\} \) (e.g., next-basket prediction).
\end{itemize}
This task type is key information and is included in $\mathcal{T}$.
Table~\ref{tab:tasks} illustrates this task categorization further with their widely used loss function and evaluation metrics.\footnote{This categorization originated from our observation of major recommendation tasks in practice. It is possible to extend it in the future as tasks evolve.}

For a task descriptor $\mathcal{T}$, 
DTIRS selects an appropriate model structure and loss function. 
Formally, the model selection function is:

\begin{equation}
    f_S = \arg\min_{f \in \mathcal{F}_\mathcal{T}} \mathbb{E} [ L(f(X'), Y) ]
\end{equation}
where:
\begin{itemize}
    \item $\mathcal{F}_\mathcal{T}$ is the set of candidate models compatible with $\mathcal{T}$, and
    \item $L(\cdot) \in \mathcal{L}_\mathcal{T}$ is the loss function compatible with $\mathcal{T}$.
\end{itemize}

\subsection{Automated Model Selection and Optimization}
DTIRS automates both model selection and hyperparameter 
tuning~\cite{bergstra2015hyperopt, kotthoff2017auto, liaw2018tune, malkomes2016bayesian, schaffer1993selecting, yang2020hyperparameter}. 
Given a dataset schema $S$, 
DTIRS determines the optimal model and hyperparameters via:

\begin{equation}
    f^*_{\mathcal{S}}, \theta^* = \arg\min_{f, \theta} \mathbb{E} [ L(f(X'; \theta), Y) ]
\end{equation}

\noindent
where $f^*_{\mathcal{S}}$ is the best model for $S$ and $\theta^*$ are the optimized parameters.





\section{Dataset Description Language (DsDL)}
\label{sec:dsdl}
The Dataset Description Language (DsDL) is a structured schema designed to describe datasets and their associated recommendation tasks in a machine-readable format. By formalizing dataset properties and task definitions, DsDL enables the Dataset- and Task-Independent Recommender System (DTIRS) to dynamically configure models without requiring manual reconfiguration. 




\subsection{Formal Specification}
DsDL represents a dataset $\mathcal{D}$ as a structured schema that closely resembles $S$ with slight modification for engineering clarity
\begin{equation}
    S' = (C, \mathcal{T})
\end{equation}
where:
\begin{itemize}
    \item $C$ lists the dataset columns, including $X$ and $Y$, and
    \item $\mathcal{T}$ describes the recommendation task(s); it is the \emph{task descriptor} and contain the description of $Y$.
\end{itemize}

The formal grammar of DsDL is expressed using an Extended Backus-Naur Form (EBNF) notation
in Listing~\ref{lst:dsdl_grammar}.

\begin{lstlisting}[basicstyle=\ttfamily\footnotesize, frame=single, caption={EBNF Grammar for DsDL}, label={lst:dsdl_grammar}]
DsDL ::= "columns" ":" "[" ColumnList "]" 
         [TimestampCol] 
         "target" ":" "[" TargetList "]"

ColumnList ::= Column { "," Column }

Column ::= "{" "col_name" ":" String "," 
               "type" ":" ColumnType "}"

ColumnType ::= "numeric" | "binary" |
               "categorical" | "ordinal" |
               "textual" | "url" |
               "list_of_numeric" | "list_of_binary" |
               "list_of_categorical" | "list_of_url" |
               "list_of_ordinal" | "list_of_textual" 

TimestampCol ::= "timestamp_col" ":" String

TargetList ::= "{" "type" ":" TargetType "," 
                   "label_col" ":" String ","
                   "key_col" ":" String "," 
                   [ "," "list_size" ":" PositiveInt 
                     "," "relevance_col" ":" String ] "}"

TargetType ::= "binary" | "numeric" |
               "ordered_list" | "unordered_list" 

String ::= <any string>
PositiveInt ::= <any positive integer>
\end{lstlisting}

This grammar defines the core components of DsDL:
\begin{itemize}
    \item \texttt{Columns}: contains the set of the dataset columns with their column names and types.
    \item \texttt{Target}: defines the prediction task, including 
    the label or target variable (\texttt{label\_col}), 
    the key or identifier for that label (\texttt{key\_col}), and 
    additional parameters for list-based tasks (\texttt{list\_size}, \texttt{relevance\_col}). 
    Without loss of generality, we require the type of \texttt{relevance\_col} to be \texttt{binary} or \texttt{numeric}, 
    where higher values means more relevant.
    \item \texttt{TimestampCol} (optional): specifies the time-related field name, if applicable. 
\end{itemize}

\subsection{Examples of DsDL for Various Recommendation Tasks}
Below are four examples of DsDL for widely known
recommender systems related tasks, 
click-through rate prediction, rating prediction, 
next-basket prediction, and top-n (ranked) item recommendation.

\subsubsection{Example: Binary Prediction for Ad Click Prediction}
The DsDL specification in Listing~\ref{lst:ex-binary} 
describes an example dataset for advertisement recommendation, 
where the task is to predict whether a user will click on an advertisement.
The dataset contains \texttt{index\_id}, \texttt{user\_id}, \texttt{ad\_id}, 
\texttt{device\_type}, \texttt{timestamp}, and \texttt{clicked}.
The model then need to perform a binary classification to predict 
whether the user will click on an ad (\texttt{clicked}).
The column (\texttt{index\_id}) serves as the unique key for each interaction.

\begin{lstlisting}[basicstyle=\ttfamily\footnotesize, frame=single, caption={Clicked Advertisement Example}, label={lst:ex-binary}]
DsDL:
    columns: [{col_name: index_id, type: numeric},
              {col_name: user_id, type: categorical}, 
              {col_name: ad_id, type: categorical}, 
              {col_name: device_type, type: categorical}, 
              {col_name: timestamp, type: numeric},
              {col_name: clicked, type: binary}]
    timestamp_col: timestamp
    target: [{type: binary, 
              label_col: clicked, 
              key_col: index_id}]
\end{lstlisting}


\subsubsection{Example: Numeric Prediction for Rating Prediction}
The DsDL specification in Listing~\ref{lst:ex-numeric} 
describes an example dataset for rating prediction, 
where the system predicts a real-valued (numerical) score 
(i.e., rating prediction in a movie recommendation system). 
The dataset includes \texttt{index\_id},
\texttt{user\_id}, \texttt{movie\_id}, \texttt{movie\_genre}, 
\texttt{user\_age}, and \texttt{rating}.
The goal is to predict the user's \texttt{rating} for the movie.

\begin{lstlisting}[basicstyle=\ttfamily\footnotesize, frame=single, caption={Movie Rating Example}, label={lst:ex-numeric}]
DsDL:
    columns: [{col_name: index_id, type: numeric}, 
              {col_name: user_id, type: categorical}, 
              {col_name: movie_id, type: categorical}, 
              {col_name: movie_genre, type: categorical}, 
              {col_name: user_age, type: numeric},
              {col_name: rating, type: numeric}]
    target: [{type: numeric, 
              label_col: rating,
              key_col: index_id}]
\end{lstlisting}

\subsubsection{Example: Unordered List for Next-Basket Prediction}
The DsDL in Listing~\ref{lst:ex-unordered} describes an example dataset 
to predict a set of products that a user is likely to purchase in the next order.
The dataset includes \texttt{user\_id}, 
\texttt{product\_id}, 
\texttt{product\_category}, 
\texttt{order\_timestamp}, 
and \texttt{bought} as columns.
The model would then need to predict an \textit{unordered list} of \texttt{product\_id} that the user is likely to buy.
The relevance signal is based on past purchases (\texttt{bought}).

\begin{lstlisting}[basicstyle=\ttfamily\footnotesize, frame=single, caption={Next-Basket Example}, label={lst:ex-unordered}]
DsDL:
    columns: [{col_name: user_id, type: categorical}, 
              {col_name: product_id, type: categorical}, 
              {col_name: product_category, 
               type: categorical}, 
              {col_name: order_timestamp, type: numeric},
              {col_name: bought, type: binary}]
    timestamp_col: order_timestamp
    target: [{type: unordered_list, 
              label_col: product_id, 
              key_col: user_id, 
              list_size: 10,
              relevance_col: bought}]
\end{lstlisting}

\subsubsection{Example: Ordered List for Top-N Movie Recommendation}
The DsDL specification in Listing~\ref{lst:ex-ordered} 
describes an example dataset for movie recommendation, 
where we need to predict a ranked list of movies for each user.
The dataset contains \texttt{user\_id}, \texttt{movie\_id}, \texttt{genre}, and \texttt{rating} columns.
The task is an \textit{ordered list} recommendation, where the model predicts the top-10 movies for each \texttt{user\_id}.
And the list is ranked based on \texttt{rating} (relevance column).

\begin{lstlisting}[basicstyle=\ttfamily\footnotesize, frame=single, caption={Movie Recommendation Example}, label={lst:ex-ordered}]
DsDL:
    columns: [{col_name: user_id, type: categorical}, 
              {col_name: movie_id, type: categorical}, 
              {col_name: genre, type: categorical}, 
              {col_name: rating, type: numeric}]
    target: [{type: ordered_list, 
              label_col: movie_id, 
              key_col: user_id, 
              list_size: 10, 
              relevance_col: rating}]
\end{lstlisting}


\subsection{Integration of DsDL with DTIRS}
DsDL serves as the foundation for dataset- and task-independent learning in DTIRS. 
Given a new dataset, DTIRS:
\begin{enumerate}
    \item Parses the DsDL schema to extract dataset properties and task definitions.
    \item Maps the \texttt{label\_col} and \texttt{key\_col} to the appropriate learning objective.
    \item Selects the best model architecture based on the \texttt{TargetType}.
    \item Configures hyperparameter tuning and feature transformations accordingly.
\end{enumerate}



\section{Proposed Levels of Automation}
\label{sec:levels}
Recommender systems typically requires 
significant human intervention for feature engineering, model selection, and hyperparameter tuning. 
Advancements in automated machine learning (AutoML), however, show promises for pathways toward fully autonomous systems~\cite{chen2024comprehensive, li2023automlp, wang2024autosr, zheng2023automl}. 
In this section, we define different levels of automation in recommendation systems 
and position Dataset- and Task-Independent Recommender Systems (DTIRS) within this framework.

\begin{table*}[ht]
\centering
\caption{Our Proposed Levels of Automation in Recommender Systems}
\small{
\begin{tabular}{m{3cm} m{3cm} m{3.5cm} m{3cm}}
\toprule
\textbf{Level} & \textbf{Feature Engineering} & \textbf{Task Generalization} & \textbf{Model Selection} \\ 
\midrule
\midrule
\textbf{Level-0} (Manual) & Fully manual & Task-specific & Manual selection  \\
\textbf{Level-1} (Dataset Independent, Task Specific) & Automated based on DsDL & Task-specific & Automatic model selection per dataset  \\
\textbf{Level-2} (DTIRS) & Automated based on DsDL & Handles various task types defined in DsDL & Adapt model architectures based on task type \\
\textbf{Beyond Level-2} & Fully autonomous & Universal recommender system foundational model & Self-learning across tasks  \\
\midrule
\bottomrule
\end{tabular}
}
\label{tab:levels}
\end{table*}

\subsection{Level-0: Manual Recommender Systems}

The majority of early recommender systems were built with full manual intervention. At this level, practitioners should:
\begin{itemize}
    \item manually select features for each dataset,
    \item experiment with different recommendation models to find the best fit,
    \item tune hyperparameters using grid search or heuristics, and
    \item implement task-specific objective functions and loss functions.
\end{itemize}
This level of automation is labor-intensive, requiring extensive domain expertise. 
Moreover, small changes in dataset composition may lead to different model performances,
and all those steps might have to be repeated to achieve superior performance.

\subsection{Level-1: Dataset-Independent, Task-Specific}
DTIRS at Level-1 introduces \emph{dataset independence} but remains \emph{task-specific}. 
This means that the same code can generalize across different datasets, 
but it is only capable of solving a single recommendation task type.

\vspace{0.2cm}
\noindent
\textbf{Key characteristics:}
\begin{itemize}
    \item \emph{Dataset-Independent}: The system can ingest new datasets via DsDL without requiring code modifications.
    \item \emph{Task-Specific}: The implementation is locked to one specific task type (e.g., Top-N ranking or CTR prediction).
    \item \emph{Automated Feature Engineering and Selection}: Features are processed based on schema properties.
    \item \emph{Automated Model Selection and Hyperparameter Tuning}: The best model is selected based on dataset characteristics within the scope of the task type it is capable of solving.
\end{itemize}

\noindent
\textbf{Limitations:}
\begin{itemize}
    \item Cannot generalize across different recommendation tasks (e.g., the pipeline built for CTR prediction cannot be used for next-basket prediction without rewriting the code).
\end{itemize}

\subsection{Level-2: Dataset- and Task-Independent}
Level-2 systems expands on Level-1 by making the system capable of handling various recommendation task types 
without modifying the core codebase. 
While it still relies on DsDL for dataset specification, 
it can automatically configure itself to handle different task types; it is a complete DTIRS.

\vspace{0.2cm}
\noindent
\textbf{Key characteristics:}
\begin{itemize}
    \item \emph{Dataset-Independent}: Works across different datasets with no manual feature selection.
    \item \emph{Task-Independent}: Can solve any recommendation task defined in DsDL.
    \item \emph{Task-Aware Model Selection}: The system dynamically selects an appropriate model based on the specified task type.
    \item \emph{Self-Configuring Pipelines}: The system adjusts preprocessing, feature engineering, loss functions, and evaluation metrics depending on the task.
\end{itemize}

\noindent
\textbf{Challenges in Achieving Level-2:}
\begin{itemize}
    \item \emph{Task Recognition}: The system must properly interpret different task types from DsDL and configure itself accordingly.
    \item \emph{Task-Specific Optimization}: Loss functions, model architectures, and hyperparameters may need distinct optimizations per task.
    \item \emph{Evaluation Metrics}: Metrics differ across tasks (e.g., NDCG for ranking vs. ROC-AUC for binary classification).
\end{itemize}

\subsection{Beyond Level-2: Fully Autonomous Recommender Systems}
While Level-2 DTIRS introduces task flexibility, there might be a possibility 
where we have 
entirely \emph{self-sufficient} recommender system 
that requires zero manual intervention (not even a DsDL). 
Such a system would:
\begin{itemize}
    \item identify and preprocess datasets without human input,
    \item determine recommendation tasks without predefined schemas,
    \item train on multiple datasets simultaneously, extracting shared knowledge, and 
    \item optimize recommendations based on feedback without explicit retraining cycles.
\end{itemize}

This represents the \emph{foundational model} paradigm for recommender systems, 
where a single model can adapt dynamically to any dataset and recommendation objective.
A concrete way to achieve this phase, however, is beyond the scope of this paper at the moment.
Table~\ref{tab:levels} summarizes our proposed levels of automation for Recommender Systems.


\section{Research Roadmap: From Level-1 to Level-2 Automation}
\label{sec:roadmap}
Transitioning from \emph{Level-1} (dataset-independent, task-specific) to 
\emph{Level-2} (dataset- and task-independent) 
requires advancements in automation, generalization, and task adaptation. This section outlines a research roadmap that defines key phases in achieving this transition.

\subsection{Phase 1: Establishing Dataset-Independent Recommendation Systems}
The first phase focuses on ensuring that the recommender system can generalize across datasets without requiring manual reconfiguration; it is the foundation of Level-1.

\vspace{0.2cm}
\noindent
\textbf{Key Challenges:}
\begin{itemize}
    \item Developing feature engineering pipelines that dynamically adapt to different dataset schemas.
    \item Automating model selection and hyperparameter tuning based on dataset characteristics.
    \item Ensuring that performance remains competitive compared to manually tuned models.
\end{itemize}

\noindent
\textbf{Milestones:}
\begin{itemize}
    \item Implementing the Dataset Description Language (DsDL) for schema standardization.
    \item Creating a benchmark suite to evaluate dataset-agnostic performance.
    \item Validating DTIRS on multiple real-world datasets.
\end{itemize}

\subsection{Phase 2: Enabling Task-Independent Systems}

To transition to {Level-2}, the system must be capable of solving 
various recommendation tasks using a unified pipeline.

\vspace{0.2cm}
\noindent
\textbf{Key Challenges:}
\begin{itemize}
    \item Automatically configuring loss functions and evaluation metrics appropriate to various tasks.
    \item Designing modular architectures that allow seamless transitions between different tasks.
    \item Optimizing for efficiency, ensuring that generalization does not come at a high computational cost.
\end{itemize}

\noindent
\textbf{Milestones:}
\begin{itemize}
    \item Understand the task at hand through the task metadata defined in DsDL.
    \item Implementing task-aware model selection and training pipelines.
    \item Evaluating DTIRS performance on benchmark datasets for various tasks.
\end{itemize}

\noindent
Finally, please note that a concrete roadmap beyond Level-2
is beyond the scope of this paper at the moment.

\section{Limitations and Open Challenges}
While DTIRS introduces significant advancements in the automation of recommender systems, several challenges remain. 
This section discusses key limitations and outlines open research problems that need further investigation.

\subsection{Generalization vs. Specialization Trade-offs}
\label{sec:genspec}
One of the fundamental challenges in designing dataset- and task-independent recommender systems is balancing
\emph{generalization} and \emph{specialization}. 
Traditional recommender systems are highly optimized for specific datasets and tasks, often achieving state-of-the-art performance through fine-tuned architectures.
On the other hand, DTIRS prioritizes reusability and automation, which may come at the cost of task-specific optimizations.

The key distinction, however, is that the DTIRS' goal is not to \emph{raise the ceiling} of performance. 
Instead, DTIRS focuses on \emph{raising the floor}.
By \emph{raising the floor}, we mean improving the baseline performance of recommender systems in a way that is universally applicable, highly reusable, and adaptable to various tasks and datasets.
Traditional systems may outperform DTIRS in specific cases, but they often require significant expertise to fine-tune for particular datasets. 
It is still an open question how DTIRS can balance task-agnostic generalization while maintaining specialized performance on individual datasets.
%

\subsection{Flat Table Dataset Representation in DsDL}
The current version of DsDL is designed to handle single-file (flat table) datasets, meaning it does not natively support multi-table or relational datasets.
Please note that, any multi-file or relational dataset can be transformed into a single-file representation by performing joins and denormalization prior to training.
While this transformation is feasible, we acknowledge that it may lead to larger dataset sizes and computational costs.
The decision to adopt a single-file approach was made to streamline the feature engineering and model development process, ensuring a uniform structure that simplifies automation.

\subsection{Computational Overhead from Automation}
DTIRS relies on automated feature engineering, 
model selection, and hyperparameter tuning, 
which introduces additional computational costs.
Traditional recommender systems are manually configured but computationally efficient once trained.
DTIRS, especially in its Level-2 automation, 
must dynamically evaluate different models and configurations, leading to higher initial computation costs.
Automated pipelines require more memory and processing power, 
making them more expensive in resource usage.
Therefore, optimizing computational efficiency through techniques such as adaptive pipeline selection, model compression, and caching strategies is an open area of research.

\subsection{Scalability for Large-Scale Deployments}
While DTIRS reduces the need for manual tuning, 
scaling it for industry-scale datasets introduces additional challenges.
Real-world datasets are highly heterogeneous and require efficient pipeline execution.
In addition, large-scale recommendation scenarios (e.g., personalized e-commerce platforms) require fast inference and real-time updates, which may be difficult for automated pipelines.
Thus, the balance between automation and real-time performance remains an open problem.

\subsection{Integration with Domain-Specific Knowledge}
Traditional recommender systems often integrate domain-specific knowledge 
(e.g., business rules, expert heuristics). 
DTIRS currently lacks a mechanism for incorporating such constraints.
This consideration raises questions, such as:
\begin{itemize}
    \item how can domain experts override automated feature selection when necessary?
    \item can DTIRS integrate external business logic without compromising automation?
    \item should there be a manual intervention layer within an otherwise automated pipeline?
\end{itemize}





\begin{table*}[ht]
\centering
\caption{Comparison of DTIRS and Traditional Recommender Systems}
\label{tab:comparison}
\small{
\begin{tabular}{m{3cm} m{4.9cm} m{4.7cm}}
\toprule
\textbf{Aspect} & \textbf{Traditional Recommender Systems} & \textbf{DTIRS} \\
\midrule
\midrule
\textbf{Adaptability} & Requires significant manual intervention for each new dataset and task, often tailored for specific use cases. & Automatically adapts to various datasets and tasks using the Dataset Description Language (DsDL) without re-engineering. \\ \hline
\textbf{Human Intervention and Expertise} & Requires extensive domain knowledge for feature engineering, model selection, and hyperparameter tuning. & Minimizes human intervention; automates feature engineering, model selection, and hyperparameter tuning, making it accessible to non-experts. \\ \hline
\textbf{Reusability} & Reduced code reusability due to dataset- and task-specific designs. Significant modifications are needed to adapt to different datasets and tasks. & High code reusability enabled by DsDL, allowing the same codebase to work across datasets and tasks. \\ \hline
\textbf{Dataset- and Task-Specific Optimization} & Capable of deep customization for specific datasets, allowing for highly optimized performance. & Trades off deep customization for generalizability, potentially leading to suboptimal performance in highly specialized tasks. \\ \hline
\textbf{Computational Overhead} & Computationally efficient due to task-specific optimizations and manual configuration focusing on relevant features and models. & Can have significant computational overhead due to automated feature engineering, model selection, and hyperparameter tuning, especially for large-scale datasets. \\ 
\midrule
\bottomrule
\end{tabular}
}
\end{table*}

\section{Conclusion}
In this paper, we have introduced 
\emph{Dataset- and Task-Independent Recommender Systems (DTIRS)}, 
a novel framework designed to eliminate the need for dataset- and task-specific configurations. 
DTIRS offers a generalized approach to recommender systems that can be easily adapted to 
new datasets and tasks without requiring manual reconfiguration.

Traditional recommender systems often require extensive manual intervention to adapt to different datasets and tasks. 
DTIRS automates these processes, enabling a more flexible and efficient approach.
%
The automation and generalization offered by DTIRS lead to improved reusability, greater accessibility, and enhanced reproducibility. 
DTIRS makes it easier for practitioners to apply recommender systems across a variety of domains, 
reducing the time and expertise required to deploy high-quality models. 
Furthermore, by providing a universal baseline, DTIRS allows researchers to focus on 
advancing the state of the art rather than reconfiguring models for each new use case.
Table~\ref{tab:comparison} summarizes the differences between DTIRS and traditional recommender system.

We encourage future research to build upon the DTIRS framework, 
addressing the challenges related to 
dataset- and task-specific performance, 
and computational overhead.
Further exploration in meta-learning, self-supervision, and cross-task optimization 
would also be essential to fully realize the potential of dataset- and task-independent recommender systems. 
By improving the core principles of DTIRS, we can create a more reusable and accessible foundation for recommender system development.



\appendix
\section{Appendix: Frequently Asked Questions}
The concept of DTIRS represents a paradigm shift, and as with any new approach, 
it raises many questions. 
Here, we address some of the most common ones.

\vspace{0.2cm}
\noindent
\textbf{Q1}: 
\emph{DTIRS may not have the benefit of tailored feature engineering like traditional recommender systems, 
which are dataset-dependent. 
It seems unlikely that DTIRS could outperform traditional methods. 
Why do we need to care about DTIRS?}

\vspace{0.1cm}
\noindent
\textbf{A1}: It is correct that DTIRS is more generalized, which means it may sacrifice some of the task-specific advantages that traditional, dataset-tailored recommender systems offer (see also Section~\ref{sec:genspec}). 
However, the key distinction is that the goal of DTIRS is not to \emph{raise the ceiling} of performance (as many traditional systems and papers aim to do). Instead, DTIRS focuses on \emph{raising the floor}.
By \emph{raising the floor}, we mean improving the baseline performance of recommender systems in a way that is universally applicable, highly reusable, and adaptable to various tasks and datasets.
Traditional systems may outperform DTIRS in specific cases, but they often require significant expertise to fine-tune for particular datasets. 

The benefit of DTIRS lies in its reusability; once a robust, generalized system is developed, it becomes
easier for practitioners to deploy across diverse datasets without requiring the same level of fine-tuning. 
In this sense, DTIRS has the potential to benefit a broader audience, providing a reliable baseline that does not require constant adjustments.

We acknowledge that \emph{ceiling-raiser} algorithms will continue to be popular in the literature, which is perfectly fine. 
What we propose is that, alongside these highly specialized advancements, 
there should also be investment in \emph{raising the floor}—the universal baseline. 
A solid baseline means that when researchers need to start with a recommendation model, they can rely on a trusted, ready-to-use system without spending unnecessary time on tweaking and feature engineering. 
This foundational work will allow researchers to focus their efforts on improving the state-of-the-art (the \emph{ceiling}) while ensuring they have a solid, reliable baseline to build upon.

If developed well, DTIRS could even serve as a global benchmark for recommendation systems, providing a standardized starting point for the community. Our goal is to shift the focus from simply raising the ceiling to also improving the universal floor.
To promote DsDL as a standard and to ensure ongoing updates, we have created an online website at \url{https://dtirs.gitlab.io} (work in progress), which will act as the central hub for all related resources.

\vspace{0.2cm}
\noindent
\textbf{Q2}: 
\emph{This paper lacks experimental results. What is the justification for the lack of experiments?}

\vspace{0.1cm}
\noindent
\textbf{A2}: 
Our primary focus is to present a paradigm shift—an approach to building recommender systems that emphasizes adaptability and generalization across tasks and datasets. 
Sections~\ref{sec:universal} and \ref{sec:dtirs} 
outline this shift, 
while Section~\ref{sec:dsdl} introduces the Dataset Description Language (DsDL), which is central to DTIRS.

The experimental results are not the focus of this paper because the core contribution is conceptual. 
DTIRS, however, imposes no restrictions on the code that can be incorporated into its implementation.
This means that if an experiment is conducted with a traditional system, it can be replicated in DTIRS with the same results, assuming the same models and hyperparameters are used. 
We include a simple prototype to demonstrate DTIRS in action, which can also be found in \url{https://dtirs.gitlab.io}.

The goal of this paper is to lay the groundwork for future research, emphasizing the need for a standardized approach and universal baseline that can evolve over time as the field progresses.

\vspace{0.2cm}
\noindent
\textbf{Q3}: 
\emph{Several frameworks aim to automate and benchmark recommender system models, 
such as~\cite{anand2020auto, anelli2021elliot, michiels2022recpack, salah2020cornac, sedhain2015autorec, sonboli2021librec, sun2020are, vente2023introducing, wang2020autorec, zhao2021recbole, zhu2022bars}. 
Many of these frameworks include implementations of various algorithms and hyperparameter tuning. 
Haven’t they already achieved what DTIRS is proposing?}

\vspace{0.1cm}
\noindent
\textbf{A3}: 
These frameworks are valuable, and some of their components can contribute to realizing DTIRS. 
However, on their own, they have not fully accomplished what DTIRS aims to achieve. 
While they offer implementations of various algorithms, support hyperparameter tuning, and can handle multiple dataset formats, the models they produce are still tied to either a specific dataset or a particular task.
In contrast, DTIRS is designed to provide predictions across datasets and tasks without requiring manual reconfiguration (provided the appropriate DsDL). 
That said, we believe that leveraging certain components from existing frameworks could accelerate the development and implementation of DTIRS.

\bibliographystyle{plain} 
\bibliography{biblio} 


\end{document}